\documentclass[11pt,twoside]{article}
\usepackage{macro-fsut-eng}
\usepackage{graphicx}

\usepackage[T1]{fontenc} % Computer Modern (CM) fonts

\usepackage{latexsym}
\usepackage{verbatim}
\usepackage{cite}

\begin{document}

\vskip 1.0cm
\markboth{A.~Tapia, D.~Melo et al.}{Study of the chemical composition of high energy cosmic rays}
\pagestyle{myheadings}

\vspace*{0.5cm}
\title{Study of the chemical composition of high energy cosmic rays using the muon LDF of EAS between $\mathbf{10^{17.25}}$ eV and 
$\mathbf{10^{17.75}}$ eV}
\author{
A. Tapia$^{1}$,
D. Melo$^{1}$,
F. S\'anchez$^{1}$,
A. Sedoski Croce$^{1}$,
J. M. Figueira$^{1}$,
B. Garc\'ia$^{2}$,
N. Gonz\'alez$^{1}$,
M. Josebachuili$^{1}$,
D. Ravignani$^{1}$,
B. Wundheiler$^{1}$,
A. Etchegoyen$^{1}$.
}
\affil{$^1$ ITeDA (CNEA, CONICET, UNSAM), Buenos Aires, Argentina. \\
$^2$ ITeDA (CNEA, CONICET, UNSAM), Mendoza, Argentina}

\begin{abstract}
We explore the feasibility of estimating primary cosmic ray composition at high energies from the study of two parameters of 
Extensive Air Showers (EAS) at ground and underground level with Monte Carlo simulations using the new EPOS and QGSJETII hadronic models tuned with LHC data. Namely, the slope and density at a given distance of the muon lateral distribution function are analysed in this work. The power to discriminate primary masses is quantified in terms of merit factor for each parameter. The analysis considers three different primary particles (proton, iron and gamma), four different zenith angles (0$^{\circ}$, 15$^{\circ}$, 30$^{\circ}$ and 45$^{\circ}$) and primary energies of $10^{17.25}$ eV, $10^{17.50}$ eV and $10^{17.75}$ eV.
\end{abstract}

\section{Introduction}
\label{sec:introduction}
The particle lateral distribution of EAS is the key quantity for cosmic ray ground observations at energies greater than $10^{15}$ eV, from which most observables are derived. An EAS is initiated by a high energy cosmic ray interacting in the top of the atmosphere and creating a multitude of secondary particles, which arrive at ground nearly at the same time. Secondary particles are distributed over a large area perpendicular to the direction of the cosmic ray primary. The disc of secondary particles may extend over several hundred meters from the shower axis, reaching its maximum density in the center of the disc, which is called the shower core. The density distribution of particles within the shower disc can be used to derive information on the primary particle. EAS measurements at ground level are carried out using arrays of individual detectors, which take samples of the shower disc at several distances from the shower core \cite{Stanev:01}.\\

It is known that the muonic component of EAS carries information about the identity of the primary particle \cite{Supa:01}. In this work we study two parameters sensitive to primary masses: the slope of the Muon Lateral Distribution Function (MLDF) and the density of muons at a certain distance from the core. The underlying idea is that showers originated by heavy nuclei produce more muons than lighter ones (therefore the absolute muon density at fixed distance will be higher for heavy nuclei) but the MLDF will be less steep. For the presented analysis the response of any detector is not taken into account.

\vspace{-0.3cm}
\section{Monte Carlo simulations}
\label{sec:MonteCarlo}
For this work we generated a library of EAS using AIRES 2.8.4a \cite{Aires} which make use of QGSJET-II-03 \cite{QGSJET:04} as hadronic model and CORSIKA 7.3700 \cite{Corsika}  which make use of QGSJET-II-04 \cite{QGSJET:04} and EPOS-LHC \cite{EPOS-LHC} as hadronic models. In both programs, we set a statistical thinning of $10^{-6}$. For each hadronic model we consider three types of primaries (proton, iron and gamma), four zenith angles (0$^{\circ}$, 15$^{\circ}$, 30$^{\circ}$ and 45$^{\circ}$) and three energies ($10^{17.25}$ eV, $10^{17.50}$ eV and $10^{17.75}$ eV). For each energy, zenith angle and primary type a total of 120 showers were produced, considering an uniform azimuthal distribution between 0$^{\circ}$ and 360$^{\circ}$. During the EAS simulations only muons with energies above 55 MeV have been taken into account to assess the MLDF at different depths between $0$ and $2.5$ m using the same set of showers.

\vspace{-0.3cm}
\section{$\beta$ and $\rho_{\mu}(500)$ parameters}
\label{sec:beta}
Once the EAS is simulated, the muon lateral density is fitted event-by-event with a KASCADE-Grande like MLDF \cite{Kascade-Grande:01} : 
\begin{equation}\label{eq:fKascade-Grande}
 \rho_{\mu}(r)=N_{\mu}\left(\frac{r}{r_0}\right)^{-\alpha}\left(1+\frac{r}{r_0}\right)^{-\beta}\left(1+\left(\frac{r}{10r_0}\right)^2\right)^{-\gamma},
 \end{equation}
 where $r$ is core distance in the shower plane. The values of $\alpha$, $\gamma$ and $r_{0}$ are fixed in $0.75$, $3$ and $320$ m, respectively, and $N_{\mu}$ and $\beta$ are the free parameters.\\ 
The MLDF is simulated at ground with AIRES and CORSIKA and afterwards propagated underground. In the last case, two depths have been used: $1.3$ m and $2.5$ m. The
layer of soil is used as a shielding against the electromagnetic component of the EAS, allowing only the muons of energy greater than $0.52$ and $1$ GeV to arrive to  the desired depths respectively. For each event simulated, $\beta$ and $\rho_{\mu}(r=500)$ are obtained from Eq. (\ref{eq:fKascade-Grande}). For the energies considered, the distance of $500$ m is close to the distance where the fluctuations over the MLDF are minimized if the any detector response is not taken into account. However, Fig. \ref{fit_1} ({\it Right}) show a wide range in distance with similar fluctuations values.   

\vspace{-0.3cm}
\section{Muons propagation through the soil}
\label{sec:Soil}
Muons lose a fraction of their energy when they propagate through the soil, mainly due to ionization. Therefore, we assume as a first-order approximation that the energy loss is proportional to the muon track length and constant with energy. Then the energy of a muon that traveled a distance $x$ through the soil is: {\small $E_{\mu}(x)=E_{\mu 0} - \alpha\rho x$}, where {\small$E_{\mu 0}$} is the initial energy of the muon, {\small$\rho=1.8\times 10^6$} g m{\small$^{-3}$} is the density of a soil with standard rock and {\small$\alpha=1.808\times 10^{-7}$} GeV m$^{2}$ g$^{-1}$ is the fractional energy loss per grammage depth.
\begin{figure}[!htp]
  \vspace{-0.4cm}
  \centering
  \includegraphics[width=6cm, height=4cm]{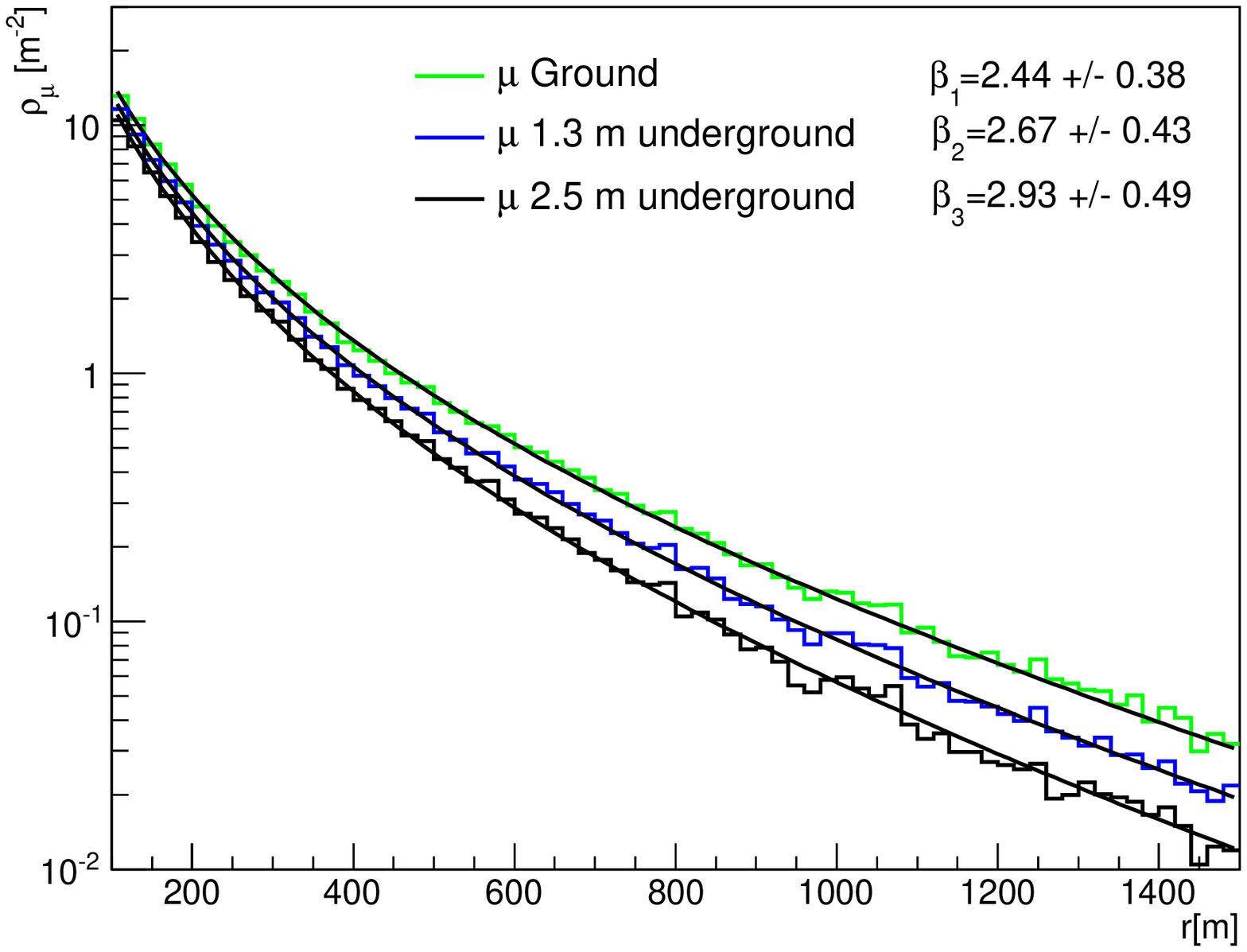} 
  \includegraphics[width=6cm, height=4cm]{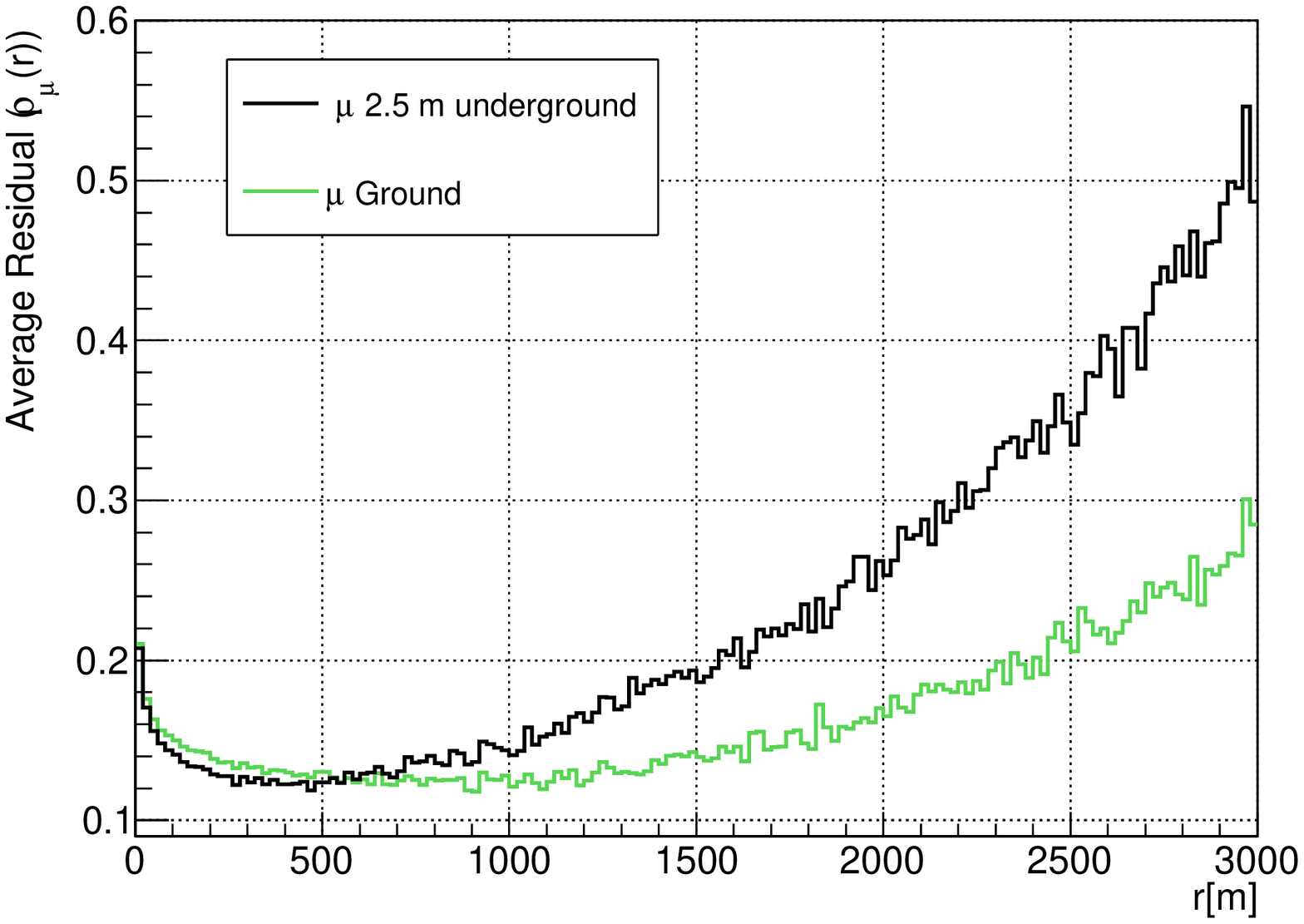}
  \caption{{\it Left:} MLDF of one vertical proton of $10^{17.75}$ eV at different soil levels. The muons propagation was performed using the first-order approximation. {\it Right:} The mean spread in  $\rho_{\mu}(r)$ for 120 showers initiated by proton with energy of $10^{17.50}$ eV and zenith angle of 30$^{\circ}$.}
    \label{fit_1}
 \vspace{-0.24cm}   
\end{figure}
As an example, in Fig. \ref{fit_1} we show a fit of the muon lateral density for one vertical proton with energy of $10^{17.75}$ eV and QGSJET-II-03 hadronic model. In this case, the slope values were {\small $\beta=2.44\pm0.38$}, {\small$\beta=2.67\pm0.43$} and {\small$\beta=2.93\pm0.49$}, and the density values were {\small$\rho_{\mu}(500)=0.81\pm 0.12$} m$^{-2}$, {\small$\rho_{\mu}(500)=0.62\pm 0.09$} m$^{-2}$, and {\small$\rho_{\mu}(500)=0.48\pm 0.07$} m$^{-2}$, for ground, $1.3$ m and $2.5$ m underground, respectively.\\
In order to validate the first-order approximation, we use the GEANT4 package to perform a complete simulation of particles arriving to different depths. The results are equivalent because the ionization is the main process of energy loss.

\section{Data Analysis}
\label{sec:DataAnalysis}
\begin{figure}[!htp]
 \vspace{-0.1cm}
  \centering
  \includegraphics[width=2.9cm, height=5cm]{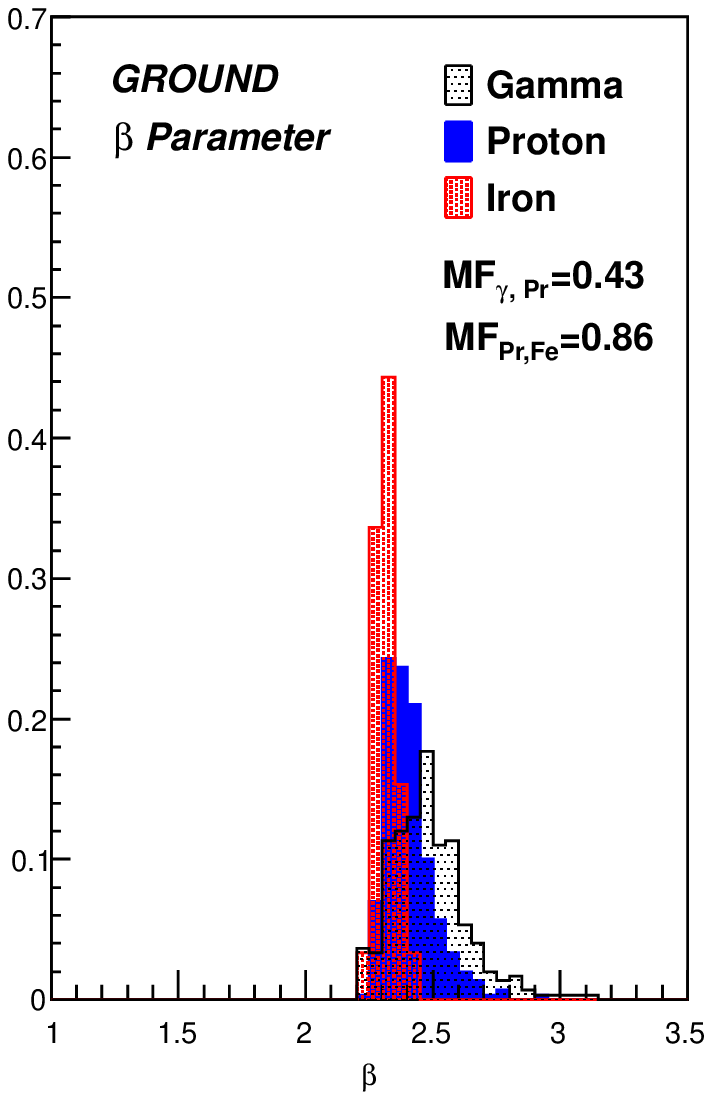} 
  \includegraphics[width=2.9cm, height=5cm]{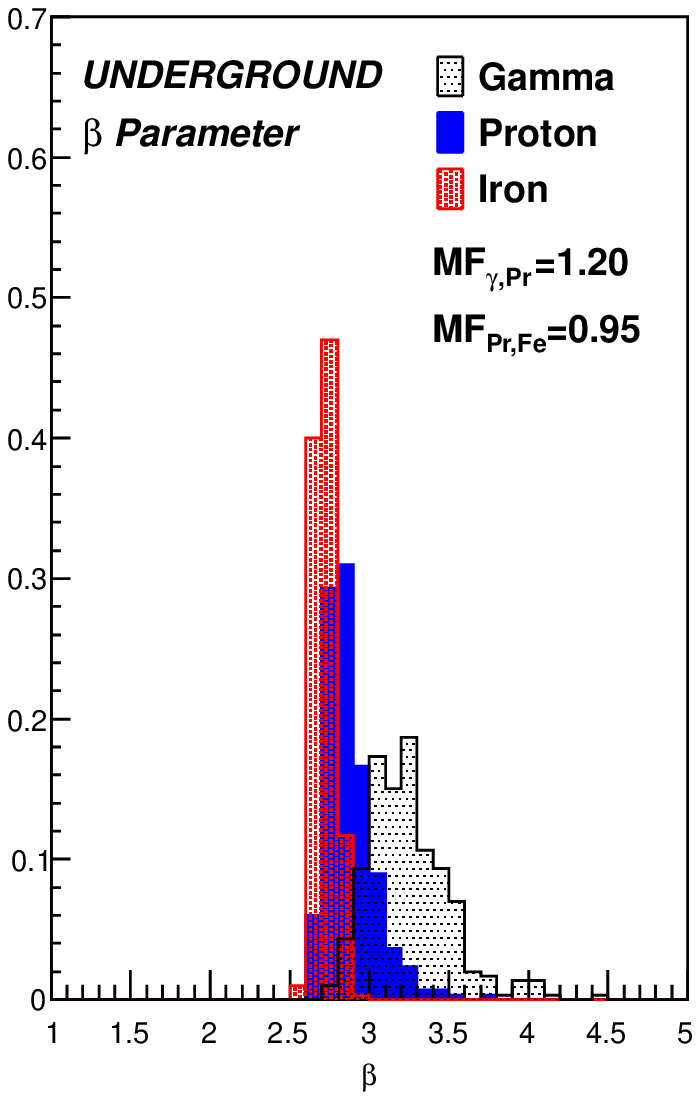}
   \hspace*{0.3cm}
  \includegraphics[width=2.9cm, height=5cm]{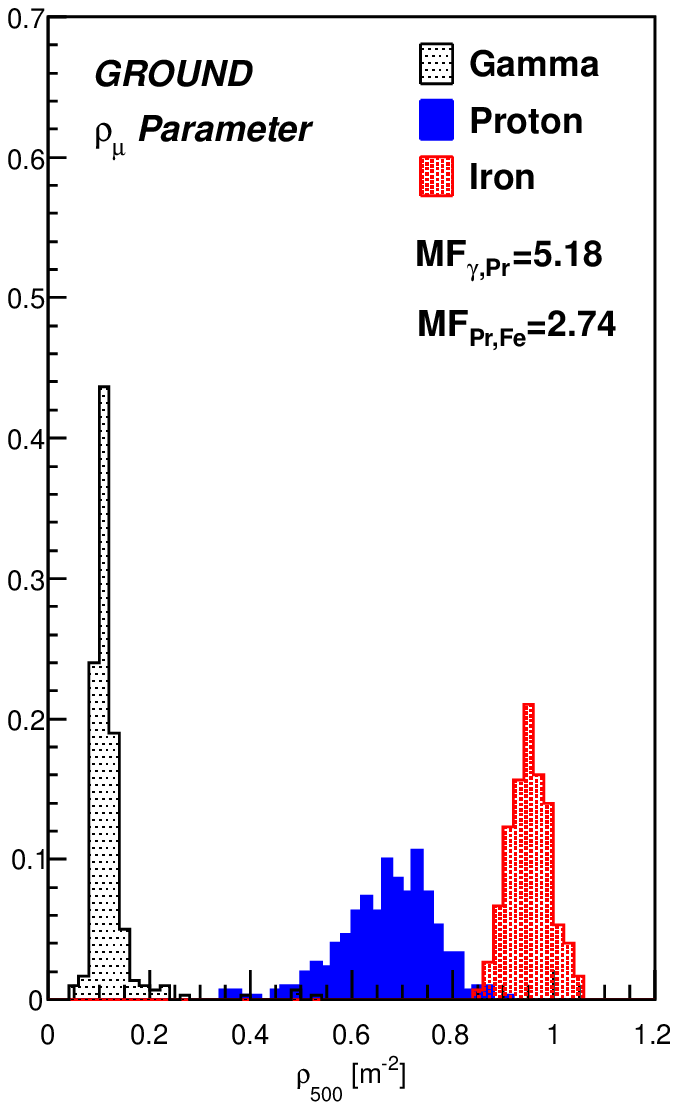}
  \includegraphics[width=2.9cm, height=5cm]{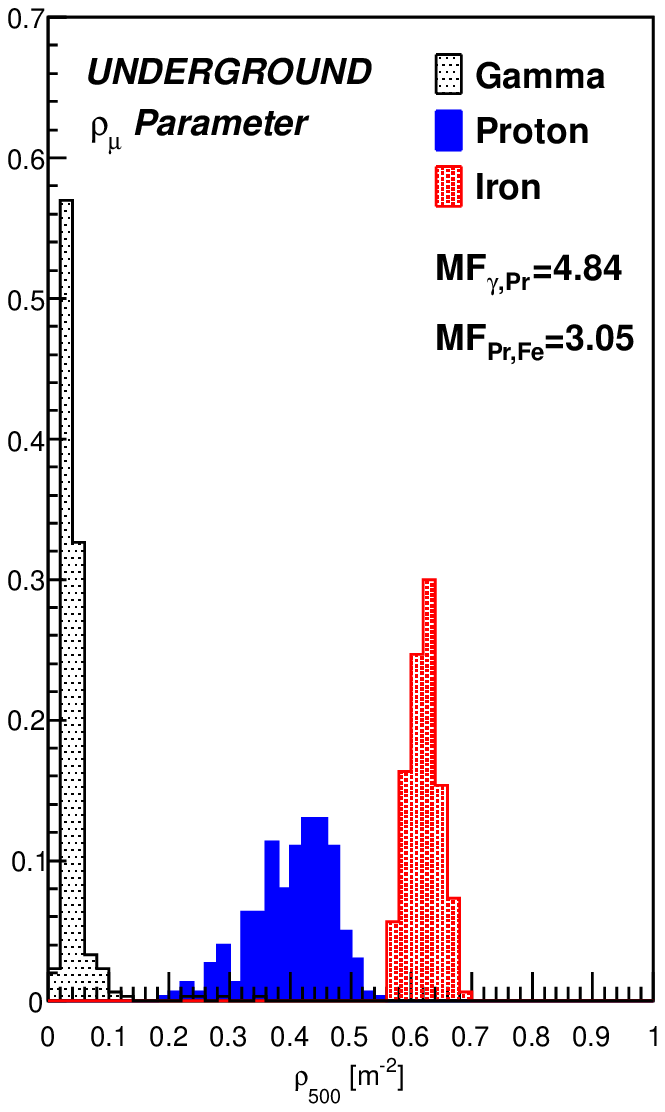}
  \caption{{\it Left:} $\beta$ parameter distributions corresponding to 120 vertical EAS at ground and $2.5$ m underground levels, with energy of $10^{17.75}$ eV and QGSJET-II-03 hadronic model. {\it Right:} The same for $\rho_{\mu}(500)$
  parameter distributions.}
    \label{beta}
  \vspace{-0.2cm}
  \end{figure}

In order to quantify the discrimination power of $\beta$ and $\rho_{\mu}(500)$ parameters, we constructed the distributions of these two parameters for each primary type, energy and zenith angle, at ground and underground levels, respectively. As an example, in Fig. \ref{beta} the distributions of $\beta$ and $\rho_{\mu}(500)$ are shown for $E=10^{17.75}$ eV and zenith angle $0^{\circ}$. The power of discrimination is measured calculating the Merit Factor (MF) between the distributions defined as: {\footnotesize                                                                                                                                                                                                                                                                                                                                                                                                                                                                                     $MF=|\langle A\rangle - \langle B\rangle|/\sqrt{\sigma_A^2+\sigma_B^2}$}, where $\langle A\rangle$ and $\langle B\rangle$ are the mean values of the $A$ and $B$ distributions, and $\sigma_A$ and $\sigma_B$  their respective standard deviations (because no detector is considered in this work, these fluctuations correspond only to shower-to-shower fluctuations). MFs lower than 1 indicate very poor discrimination power (roughly because mean values are less than $1\sigma$ apart from each other).

\vspace{-0.4cm}
\section{Merit factor and primary mass discrimination}
\label{sec:MeritFactor}
\begin{figure}[!htp]
   \vspace{-0.2cm}
  \begin{center}
   \includegraphics[width=6.5cm, height=5cm]{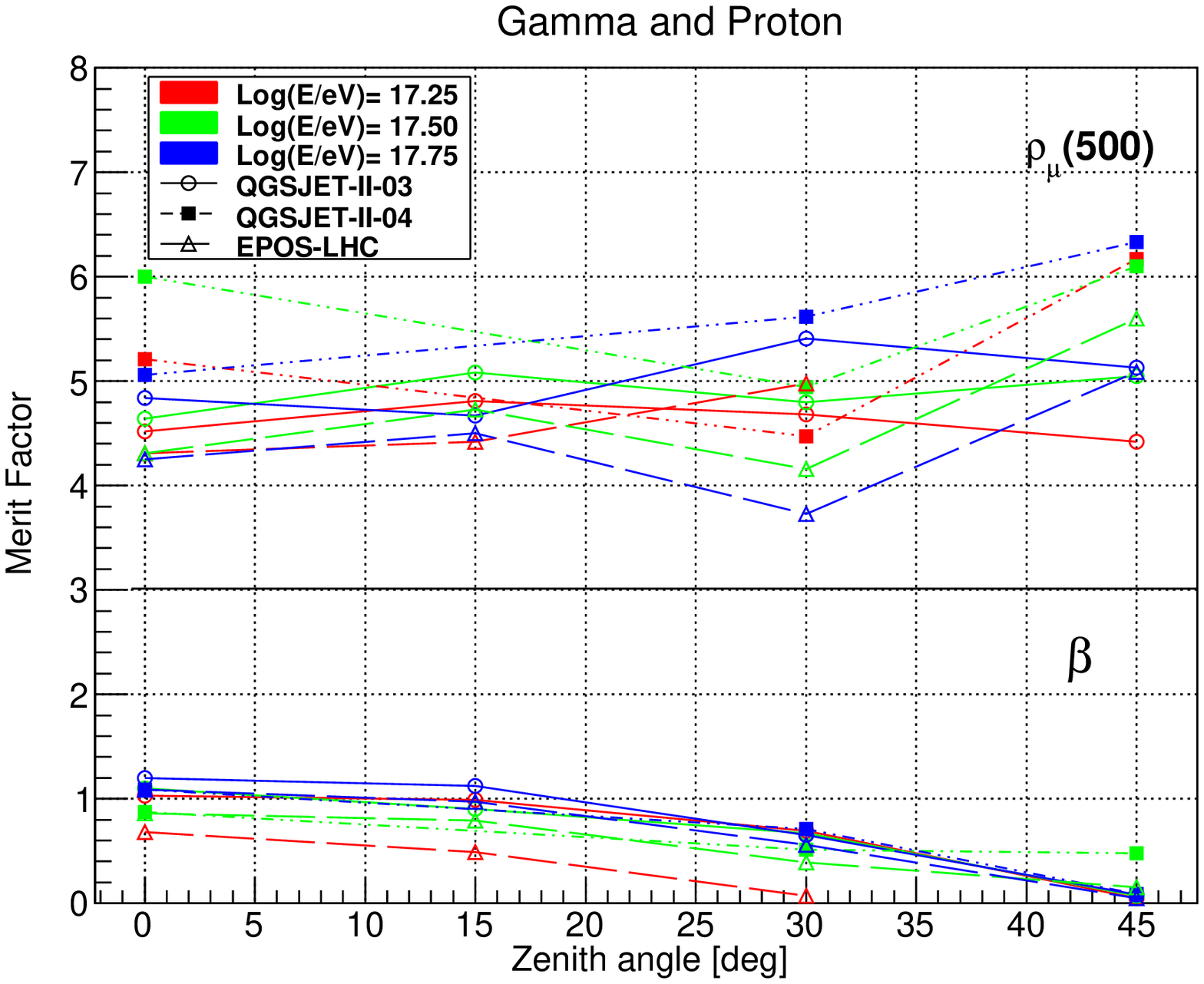} 
   \includegraphics[width=6.5cm, height=5cm]{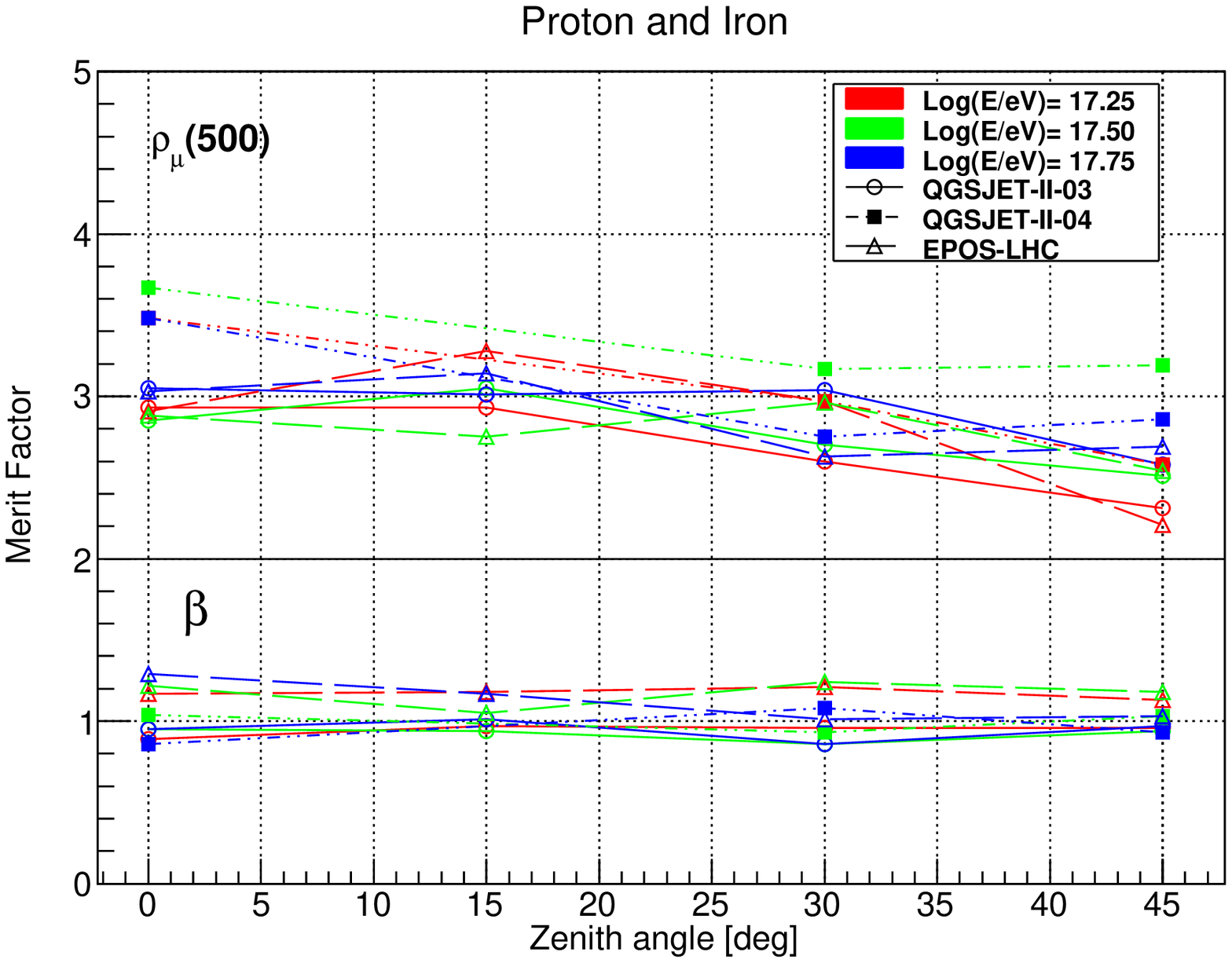}
  \caption{Merit factor of the $\rho_{\mu}(500)$ and $\beta$ parameter at $2.5$ m underground level as function of primary energy and zenith angle. {\it Left:} For Gamma and Proton. {\it Right:} For Proton and Iron.}
  \label{beta_merit_factor}
  \end{center}
 \end{figure}

The MFs dependence with energy, hadronic model and zenith angle was the main goal of this work. In Fig. \ref{beta_merit_factor} we show the results for case of $2.5$ m underground. From both plots it can be seen that the discrimination power of $\beta$ is not very good. On the other hand, $\rho_{\mu}(500)$ shows a MF greater than $2$ and is practically independent of the energy and zenith angle. Therefore this parameter has the potential to discriminate the mass of primaries cosmic rays. A similar behavior was seen at $1.3$ m underground level.

 \vspace{-1cm}
\section{Conclusions}
In this work we performed a study of the merit factor of $\beta$ and $\rho_{\mu} (500)$ parameters obtained from MLDFs. We used proton, iron and gammas as primary
cosmic ray, three hadronic models and different ground levels. The muons propagation through of soil was performed using an approximation which only took into account the continual energy loss. For energies and zenith angles studied, the $\beta$ parameter has a merit factor $\leq 1$, and therefore it is not a good mass discriminator, while the $\rho_{\mu}(500)$ parameter has a merit factor $> 2$ (if Poissonian fluctuations are not taken into account). In each case, the merit factor does not show a significant difference with the three hadronic models considered.

\end{document}